\documentclass[final,3p,times]{elsarticle}
\usepackage{lineno,hyperref}
\modulolinenumbers[5]

\journal{Physica A}

\biboptions{numbers,sort&compress}
\usepackage{rotating}
\usepackage{graphicx}
\usepackage{color}
\usepackage{caption}

\newcommand\eq[1]{Eq.~(\ref{#1})}

\newcommand{\sign}{\textup{sign}}

\begin{document}
	
\begin{frontmatter}
	
	\title{Exact solution for the order parameter profiles and the Casimir force in $^4$He superfluid films in an effective field theory}
	
	
	\author[mymainaddress,mysecondaryaddress]{Daniel Dantchev}
	\author[mysecondaryaddress]{Joseph Rudnick}
	\author[mymainaddress]{Vassil M Vassilev} 
	\author[mymainaddress]{and Peter A Djondjorov}
	
	\cortext[cor1]{Corresponding author: Daniel Dantchev}

\address[mymainaddress]{Institute of Mechanics -- Bulgarian Academy of Sciences, \\
	Acad. Georgy Bonchev St., Building 4,	1113 Sofia, Bulgaria}
\address[mysecondaryaddress]{Department of Physics and Astronomy, UCLA, Los Angeles,
	California 90095-1547, USA}	
	
	\ead{daniel@imbm.bas.bg, jrudnick@physics.ucla.edu, vasilvas@imbm.bas.bg and padjon@imbm.bas.bg}

\date{\today}

\begin{abstract}
We present an analytical solution of an effective field theory which, in one of its formulations, is equivalent to the Ginzburg's $\Psi$-theory for the behavior of the Casimir force in a film of $^4$He in equilibrium with its vapor near the superfluid transition point. We consider three versions of the theory, depending on the way one determines its parameters from the experimental measurements. We present exact results for the behavior of the order parameter profiles and of the Casimir force within this theory, which is characterized by  $d=3$, $\nu=2/3$ and $\beta=1/3$, where $d$ is the bulk spatial dimension and $\nu$ and $\beta$ are the usual critical exponents. In addition, we revisit  relevant experiments  \cite{GC99} and \cite{GSGC2006} in terms of our findings. We find reasonably good agreement between our theoretical predictions and the experimental data. We demonstrate analytically that our calculated force is attractive. The position of the extremum is predicted to be at  $x_{\rm min}=\pi$, with
 $x=(L/\xi_0)(T/T_\lambda-1)^{1/\nu}$, which value effectively coincides with the experimental finding $x_{\rm min}=3.2\pm 0.18$. Here $L$ is the thickness of the film, $T_\lambda$ is the bulk critical temperature and $\xi_0$ is the correlation length amplitude  of the system for temperature $T>T_\lambda$. The theoretically predicted position of the minimum does not depend on the one adjustable parameter, $M$, entering the theory. The situation is different with respect to the largest absolute value of the scaling function, which depends  on both $\xi_0$ and on $M$.  For this value the experiments yield $-1.30$. If one uses $\xi_0=1.63$ $\AA$, as in the original $\Psi$ theory of Ginsburg, one obtains the closest approach to the experimental value  $-1.848$ with $M=0$. If one uses $M=0.5$, as inferred from other experiments, along with the best currently accepted  experimental value for $\xi_0$, $\xi_0=1.432$ $\AA$, then the maximum value of the force is predicted to be $-1.58$. The effective theory considered here is not consistent with critical point universality; furthermore it incorrectly predicts ordering in the film in violation of known rigorous results. These issues are  discussed in the text. 
\end{abstract}


\end{frontmatter}
\section{Introduction}

\subsection{Critical Casimir force near the $\lambda$ transition in $^4$He}

It is now a well established experimental fact \cite{GC99,GSGC2006} that the thickness of a Helium film in equilibrium with its vapor decreases near and below the bulk transition into a superfluid state. The phenomenon has been discussed theoretically in a series of works; see, e.g., Refs. \cite{KD92a,KD92b,ZRK2004,W2004,DKD2005,MGD2007,ZSRKC2007,H2007,VGMD2007,VGMD2009,BBSBH2010,Ha2009,H2009,H2010,D2013,D2014,V2015,D2018}. Among the methods used are Renormalization Group techniques \cite{KD92a,KD92b,W2004,D2013,D2014,D2018}, mean-field type theories \cite{ZSRKC2007,MGD2007} and Monte Carlo calculations \cite{DKD2005,H2007,VGMD2007,VGMD2009,Ha2009,H2009,H2010,V2015}. Of note is  \cite{BBSBH2010} in which the role of fluctuations around the mean-field theory is taken into account perturbatively.  In all of the above approaches it is assumed that the microscopic molecular interactions are not altered by the transition and that the observed change in the thickness results from the  cooperative behavior of the constituents of the fluid system. Furthermore, the overall behavior of the force is in relatively good agreement with finite-size critical point scaling theory \cite{Ba83,PF84,C88,Ped90,PE90,BDT2000}.  A review of some of the above mentioned studies, as well as some other aspects of critical Casimir effect, can be found in \cite{Krech1994,BDT2000,G2009}. 

Inspection of the range of theoretical approaches used to study the Casimir force in Helium films reveals that the problem has, so far, not been studied in the context of the so-called $\Psi$-theory of Ginzburg and co-authors \cite{GS76,GS82}. This theory has been used by Ginzburg, et al., to describe a variety of phenomena observed in Helium films and represents a portion of the research on Helium for which  Ginzburg was recently awarded the Nobel prize in physics. In the current study we aim to fill that gap by applying  $\Psi$ theory to calculate the critical Casimir force of a Helium film that is subject only to short-ranged interactions, which is to say we neglect the van der Waals interaction between the film and its substrate. 

We study the Casimir force in a horizontally positioned liquid $^4$He film supported on a substrate when that film is in equilibrium with its vapor. We will do this for temperatures at, and close to, the critical temperature, $T_\lambda$, of $^4$He at its bulk phase transition from a normal to a superfluid state. 

\subsection{Background: data and experimental findings}

The continuous phase transition in $^4$He from a normal to a superfluid state, referred to as the $\lambda$ transition because of the temperature dependence of the specific heat,  occurs  at a temperature \cite{GS76} $T_\lambda=2.172$ $^\circ$K at a saturated-vapor pressure $p_\lambda =0.05$ atm and density \cite{W67,GS82} $\rho_\lambda=0.1459$ g/cm$^3$. We note that while the density changes continuously through the transition its temperature gradient varies discontinuously \cite{W67,DB98}. 

The critical exponents of systems, that belong to the $O(2)$ universality class of $O(n)$, $n\ge 2$ with continuous symmetry of the order parameter, are \cite{KS2001,ZJ2002,PV2002}
\begin{equation}
\label{eq:crit_exp}
\alpha = -0.011\pm 0.004  ,\qquad  \nu= 0.6703 \pm 0.0013, \qquad \eta =  0.0354\pm 0.0025.
\end{equation}
Since hyperscaling holds, all critical exponents can be determined from, say, $\nu$ and $\eta$ using the appropriate scaling relations \cite{BDT2000,KS2001,ZJ2002,PV2002}. 

\subsection{The Casimir force}

Finite-size scaling theory \cite{Ba83,C88,Ped90,PE90,BDT2000} for isotropic systems in which hyperscaling holds predicts a Casimir force of a system with a film geometry $\infty^{d-1}\times L$ of the form 
\begin{equation}
\label{eq:Cas_force_def}
\beta F_{\rm Cas}(T,L)=L^{-d} X_{\rm Cas}(a_t \hat{t} L^{1/\nu}).
\end{equation}
Here $X_{\rm Cas}$ is a universal scaling function that depends on the bulk and surface universality classes, $\hat{t}=(T-T_\lambda)/T_\lambda$, and $a_t$ is a nonuniversal metric factor. Helium 4 belongs to the $O(2)$ bulk universality class and the boundary conditions on a Helium film on a solid substrate that is in equilibrium with vapor are Dirichlet, in that the superfluid order parameter vanishes at the boundaries. 

The Casimir force $F_{\rm Cas}(T,L)$ can be expressed in terms of an excess pressure $P_L(T)-P_b(T)$:
\begin{equation} \label{Casimir}
\beta F_{\rm Cas}(T,L)= P_L(T)-P_b(T).
\end{equation} 
Here $P_L$ is the pressure on the finite system, while $P_b$ is the pressure in the infinite system. The Casimir force results from finite size effects, which are especially pronounced and of {\it universal} character near a critical point of the system. The above definition is equivalent to another commonly used relationship \cite{E90book,K94,BDT2000}
\begin{equation}
\label{grand_can}
\beta F_{\rm Cas}(T,L)\equiv-\frac{\partial\omega_{\rm ex}(T,L)}{\partial L}=-\frac{\partial\omega_L(T,L)}{\partial L}-P_b,
\end{equation}
where $\omega_{\rm ex}=\omega_L-L\,\omega_b$ is the excess grand potential per unit area, $\omega_L$ being the grand canonical potential of the finite system, again per unit area, and $\omega_b$ is the grand potential per unit volume of the infinite system. The equivalence between the definitions \eq{Casimir} and \eq{grand_can} arises from the observation that  $\omega_b=- P_b$, while for the finite system one has  $-\partial\omega_L(T,L)/\partial L=P_L$. 

\subsection{On the Casimir force in a class of systems}

It is possible to derive a simple expression for the Casimir force in systems in which the order parameter is found by minimizing a potential that does not explicitly depend on the coordinate perpendicular to the film surface. For purposes of notation we denote the spatial coordinate perpendicular to the substrate by $z$.  We consider systems in which the grand potential per unit area ${\cal \omega_A}$ is obtained by minimizing the functional 
\begin{equation}
\label{eq:functional}
{\cal \omega_A}=\int_{0}^{L} {\cal L}\left[ \phi(z),\dot{\phi}(z)\right] dz,
\end{equation}
where $\phi(z)$ is the local value of the order parameter at coordinate $z$, and $\dot{\phi}(z)\equiv d\phi(z)/dz$. We take $\cal L$ to be of the form
\begin{equation}
\label{eq:L}
\mathcal{L}=\frac{1}{2} \dot{\phi}^2(z)-f[\phi(z)]. 
\end{equation}
Following Gelfand and  Fomin \cite[pp. 54-56]{GF63} it is easy to show that the functional derivative of  ${\cal \omega_A}$ with respect to the independent variable $z$ at $z=L$ is 
\begin{equation}
\label{eq:funct_deriv}
-\left(\frac{\delta {\cal \omega_A} }{\delta z}\right)\Bigg|_{z=L} =-\left(\dot{\phi} \frac{\partial {\cal L}}{\partial \dot{\phi}}-{\cal L}\right)\Bigg|_{z=L}.
\end{equation}
Taking into account the physical meaning of this functional derivative and performing the requisite calculations we obtain 
\begin{equation}
\label{eq:pl_def}
P_L\equiv \left(\frac{\delta {\cal \omega_A} }{\delta z}\right)\Bigg|_{z=L} =\left(\frac{1}{2}\dot{\phi}^2+f(\phi)\right)\Bigg|_{z=L}.
\end{equation}
The extrema of the functional ${\cal \omega_A}$ are determined by the solutions of the corresponding Euler-Lagrange equation which leads to 
\begin{equation}
\label{eq:EL_eq_expl}
\frac{d}{dz} \dot{\phi}+\frac{\partial f}{\partial \phi}=0.
\end{equation}
Multiplying by $\dot{\phi}$ and integrating one obtains the corresponding first integral of the above second-order differential equation. The result is
\begin{equation}
\label{eq:first_int}
\frac{1}{2}\dot{\phi}^2+f(\phi)={\rm const}=P_{L}.
\end{equation}
Thus, the expression for $P_L$ has the same value at {\it any} point of the liquid film. 

Let us now assume that the boundary conditions are such that there is a point at which $\dot{\phi}=0$ and let $\phi_0$ be the value of $\phi$ at that point. Then we arrive at the very simple expression for the pressure on the boundaries of  the finite system
\begin{equation}
\label{eq:PL_final}
P_L=f(\phi_0). 
\end{equation}
When the system is infinite the gradient term decreases with distance from a boundary, asymptotic to zero in the bulk  within the type of theories we consider.  It is easy to verify that the bulk pressure is 
\begin{equation}
\label{eq:Pb_final}
P_b=f(\phi_b), 
\end{equation}
where $\phi_b$ is the solution of the equation $\partial f/\partial \phi=0$, for which $\omega_b=-f(\phi)$ attains its minimum. The excess pressure,  and hence the Casimir force, is 
\begin{equation}
\label{eq:Cas_gen}
F_{\rm Cas}\equiv P_L-P_b=f(\phi_0)-f(\phi_b). 
\end{equation}
The above expression, as we will see, is very convenient for the determination of the Casimir force  in a system that can be described by a functional of the type given in \eq{eq:functional}. It has previously been used for systems described by the Ginzburg-Landau-Wilson functional \cite{ZSRKC2007,DVD2016,Note1}. 

\section{The model}

We start by recalling some basic expressions from the usual description of the phase behavior of $^4$He films.

\subsection{Basic expressions for describing the phase behavior of $^4$He films}

We consider a film with thickness $L$ of liquid $^4$He that is in equilibrium with its vapor. We suppose the film to be parallel to the $(x,y)$ plane and its thickness to be along the $z$ axis. A constituent of the liquid film with total density $\rho$ is in the superfluid state with density $\rho_s(z)$ while the other one with density $\rho_n(z)$ is in the normal state. Obviously
\begin{equation}
\label{eq:rhos}
\rho(z)=\rho_n(z)+\rho_s(z).
\end{equation} 
We consider two order parameters: a one-component order parameter $\rho_n$ to represent the normal fluid and a two-component parameter $\Psi_s=\eta \exp(i \varphi)$ to stand for the superfluid portion of it. As usual, we take $\eta=\eta(z)$ and $\varphi=\varphi(z)$ to be real valued functions and, thus, $|\Psi_s|=\eta$ with the identification that
\begin{equation}
\label{eq:rhos_eta}
\rho_s=m |\Psi_s|^2=m\eta^2, 
\end{equation}
where $m$ is the mass of the Helium atom.  A spatial gradient of the phase of the $\Psi_s$ function gives rise to the superfluid velocity via the relationship
\begin{equation}
\label{eq:vs}
\vec{v}_s=\frac{\hbar}{m}  \nabla \varphi. 
\end{equation}

In the remainder of this article we consider only the case of a fluid at rest. Then one can take $\Psi_s$ to be a real function characterized solely by its amplitude $\eta$.

In terms of $\rho_s=m|\Psi_s|^2$ and $\rho_n$, the total amount of Helium atoms in the fluid (normalized per unite area) is
\begin{equation}
\label{eq:rho}
\rho \equiv \frac{1}{ L}\int_{0}^{L} \left[ \rho_s(z) + \rho_n(z)\right ] dz=\frac{1}{ L}\int_{0}^{L} \rho(z) dz,
\end{equation}
where the value of the overall average density $\rho$ is fixed by the chemical potential $\mu$. The above equation intertwines the profiles $\rho_s$ and $\rho_n$.  For $\rho_s$ the natural boundary conditions at both the substrate-fluid interface and the fluid-vapor interfaces are
\begin{equation}
\label{eq:bcpsis}
\rho_s(0)= \rho_s(L)=0 \Leftrightarrow \Psi_s(0)=\Psi_s(L)=0.
\end{equation}
The corresponding natural boundary conditions for $\rho_n$ depend on the interface. At the liquid-vapor interface one has 
\begin{equation}
\label{eq:bcpsin}
\rho_n(L)=\rho_b(T),
\end{equation}
where $\rho_b(T)$ is the bulk density of the liquid Helium at temperature $T$; at the substrate-liquid interface one has the so-called ``dead" layers. In these layers $^4$He has solid-like properties, i.e., it does not possess a properties of a liquid,  and it is immobilized at the boundary. This implies that there is some sort of close packing of the Helium atoms. The number of layers is generally small---from two well below $T_\lambda$ to the order of 10 in the vicinity of that temperature.  This can be thought of as a sort of adjusted thickness of the liquid films and will be ignored in our theory. Thus, we will assume that the boundary condition  (\ref{eq:bcpsin}) is fulfilled at the both boundaries of the system, i.e., that
\begin{equation}
\label{eq:bcpsinormal}
\rho_{n}(0)=\rho_n(L)=\rho_b(T).
\end{equation}

Since we are addressing a spatially inhomogeneous problem, its  proper treatment requires the minimization of the total thermodynamic potential ${\cal \omega_A}(\mu,T)$ \cite{GS82,So73}, 
which is normalized per unit area, simultaneously with respect to $\Psi_s(z)$ and $\rho(z)$. Hereafter the dot will mean a differentiation with respect to the coordinate $z$.

We take as our basic variables $\rho_s$ and $\rho$. We assume that they both vary within the film, so our system will depend on $\rho_s$ and $\rho$ and their gradients  $\dot{\rho_s}$ and $\dot{\rho}$. If however, the gradient of  $\rho$ is small, spatial derivatives
of $\rho$ can be neglected. For temperatures well below the liquid-vapor critical point we will take $\rho$ to be a constant within the film, i.e., $\rho$ is $z$-independent. This implies near the $\lambda$ point one can treat Helium as an incompressible liquid. This is what is done in \cite{GS76} and \cite{So73}.

For the total thermodynamic potential ${\cal\omega_A}(\mu,T)$ per unit area one has
\begin{equation}
\label{eq:funct}
{\cal\omega_A}(\mu,T) =  \int_{-L/2}^{L/2} \left[\omega(\mu,T,\rho, \Psi_s,\dot{\Psi_s}) - \mu \rho \right] dz,
\end{equation}
where $\omega(z)$ is the local density of this potential per unit area. Here $\omega=\omega_I(\mu,T, \rho)+\omega_{II}(\mu,T,\Psi_s,\dot{\Psi_s})$, where $\omega_I$ is the local potential density of the normal fluid and $\omega_{II}$ is that of the superfluid. Since $\mu, T$ and $\rho$ are constants through the thickness of the film, one concludes that the terms $\omega_I$ and $\mu \rho$ will generate only bulk-like contributions, after the integration. For this reason we will not be interested in the specifics of these terms.   Following \cite{GS82}, one can write 
\begin{equation}
\label{eq:omegaII}
\omega_{II}=\omega_{II,0}+\frac{1}{2 m}|-i\hbar\dot{\Psi}_s|^2,
\end{equation}
where $\omega_{II,0}=\omega_{II,0}(\mu,T,|\Psi_s|^2)$ captures the corresponding bulk potential density of the infinite system. The gradient term can easily be rewritten in the equivalent forms
\begin{equation}
\label{eq:eq_forms_grad_term}
|-i\hbar\dot{\Psi}_s|^2 = \frac{\hbar^2}{2m}\dot{\eta}^2+\frac{\hbar^2}{2m}\eta^2 \dot{\varphi}^2 = \frac{\hbar^2}{8m^2} \frac{\dot{\rho}_s^2}{\rho_s}+\frac{1}{2}\rho_s v_s^2.
\end{equation}
For a fluid at rest $v_s=0$. The conditions for the minimum of ${\cal\omega_A}(\mu,T)$ are given by the corresponding Euler-Lagrange equations, which read
\begin{equation}
\label{eq:EL_rho}
-\frac{\partial \omega_I}{\partial \rho}+\mu=0,
\end{equation}
and 
\begin{equation}
\label{eq:EL_psi}
\frac{d}{dz} \frac{\partial \omega_{II}}{\partial \dot{\rho}_s}-\frac{\partial \omega_{II}}{\partial \rho_s}=0.
\end{equation}
Note that the condition of $\rho$ being $z$-independent requires that the profiles $\rho_n(z)$ and $\rho_s(z)$ are connected; a change in one of them leads to a change in the other. We stress that the above arguments are \textit{not} dependent on the actual functional form of $\omega$; they rely simply on the assumption that the overall density of the fluid inside the film does not change. 

Keeping in mind \eq{eq:omegaII}  and \eq{eq:EL_psi} for the function $\Psi_s$, one obtains the equation
\begin{equation}
\label{eq:psi_eq}
\frac{\hbar^2}{2m} \ddot{\Psi}_s=\Psi_s \frac{\partial \omega_{II}}{\partial |\Psi_s|^2}.
\end{equation}

\subsection{On constructing effective field theories for $^4$He}

Let us consider a class of theories for which 
\begin{equation}
\label{eq:universality}
\beta\,\omega_{II,0}=|\tau|^{2-\alpha}f\left(\left|\frac{\Psi_s}{|\tau|^\beta\Psi_{s,e0}}\right|^2\right).
\end{equation}
Here $\alpha$ is the critical exponent of the specific heat, $\beta$ is that one of the order parameter, $\Psi_{s,e0}$ is the amplitude of the temperature dependence of
the equilibrium value of $\Psi_s$ in bulk Helium
and 
\begin{equation}
\label{eq:def_tau}
\tau=\left[T_c-T\right]/T_c,
\end{equation}
with $T_c$ being the bulk critical point, $T_{\lambda}$, in the case of superfluid $^4$He. The above structure of  $\omega_{II,0}$ guarantees the existence of proper thermodynamic scaling relations within  the envisaged theory. At large argument $x$, $\omega_{II,0}$ is a slowly varying function of 
$\tau$, and therefore $f(x)$ must have the form
$f(x)\propto x^{(2-\alpha)/\beta}$ when $x\gg 1$. For small $x$ the function $f(x)$ is approximately analytical and can be
expanded in powers of $x$. The most straightforward approach based on the above considerations leads to  a polynomial approximation to \eq{eq:universality}. Then $1/\beta$ is an integer. The simplest realizations are {\it i)} $\alpha=0$ and $\beta=1/2$, or {\it ii)} $\alpha=0$ and $\beta=1/3$, etc. Requiring that hyperscaling is valid, i.e., $d\nu=2-\alpha$, we conclude that $\nu/\beta$ is also an integer. When $d=4$ one is led to case {\it i)} with $\nu=\beta=1/2$, and {\it ii)} when $d=3$ to $\nu=2/3$ and $\beta=1/3$. One readily recognizes in case {\it i)} mean-field theory, while in case {\it ii)} one is led to  the so-called $\Psi$-theory of Ginzburg---see \ref{ap:psi_theory} and the text below. Since the critical exponents $\alpha=0, \nu=2/3$ and $\beta=1/3$ are numerically quite close to those of $^4$He, the effective theory represents an approximation to the known scaling properties of $^4$He, including the Casimir force in $^4$He films for $d=3$. In this case, in the absence  of an ordering field that couples linearly to the order parameter (and thus violates gauge invariance) one has
\begin{equation}
\label{eq:f_x_expansion}
f(x)\simeq \left\{ \begin{array}{cc}
A^+ (-x^2+A_4x^4+A_6 x^6),& \tau>0\\
A^- (x^2+A_4x^4+A_6 x^6),& \tau<0
\end{array}. \right.
\end{equation}
The corresponding equations for the behavior of $x$, which provides the minimum of the above expression for $f(x)$ are 
\begin{equation}
\label{eq:eq_for_x}
\frac{d}{dx}f(x)=0 \Leftrightarrow \left\{ \begin{array}{cc}
x(-x+2A_4x^2+3A_6 x^4)=0,& \tau>0\\
x(x+2A_4x^2+3A_6 x^4)=0,& \tau<0
\end{array}. \right.
\end{equation}
Requiring $x=1$ to be the solution for $\tau>0$ we obtain, after  replacing  $A_6$ with $M/3$, the result that $A_4=(1-M)/2$. Explicitly, one has
\begin{equation}
\label{eq:f_x_expansion}
f(x)\simeq \left\{ \begin{array}{cc}
A^+ \left(-x^2+(1-M)/2\; x^4+M/3\; x^6\right),& \tau>0\\
A^- \left(x^2+(1-M)/2\; x^4+M/3 \;x^6\right),& \tau<0
\end{array}, \right.
\end{equation}
where $0\le M\le 1$ ensures that $x=0$ is the only real solution for $\tau<0$. For a finite system, taking into account that the free energy has no singularity at $\tau=0$, one concludes that $A^+=A^-=A$.   In terms of the variable $\phi$, where 
\begin{equation}
\label{eq:phi_def}
\phi=\left|\frac{\Psi_s}{|\tau|^{\beta}\Psi_{s,e0}}\right|,
\end{equation}
\eq{eq:psi_eq} reads
\begin{equation}
\label{eq:psi_red_eq}
\frac{\hbar^2}{2m} \Psi_{s,e0}^2 \,|\tau|^{2\beta}\,  \ddot{\phi}=\phi \frac{\partial \omega_{II}}{\partial |\phi|^2}.
\end{equation}
Using \eq{eq:psi_red_eq}, for the spatial behavior of $\phi$ one obtains 
\begin{equation}
\label{eq:phi_spacial_eq}
\frac{\hbar^2}{2m} \Psi_{s,e0}^2 \,|\tau|^{-4/3}\,  \ddot{\phi}=(A/\beta)\, \phi \left[-\sign(\tau) + (1-M) \left|\phi\right|^2 + M\left|\phi\right|^4 \right].
\end{equation}
Introducing 
\begin{equation}
\label{eq:_corr_lenght}
\xi_\tau=\xi_0|\tau|^{-\nu}, \qquad \mbox{with} \qquad \nu=2/3, \qquad \mbox{and} \qquad \xi_0\simeq\frac{\hbar \Psi_{s,e0} }{\sqrt{2m A k_B T_c}},
\end{equation}
the above equation takes the form 
\begin{equation}
\label{eq:phi_final}
\ddot{\phi}=\phi \left[-\sign(\tau) + (1-M) \left|\phi\right|^2 + M\left|\phi\right|^4 \right],
\end{equation}
where the derivative is taken with respect to $\zeta_\tau=z/\xi_\tau$.

From \eq{eq:universality} and \eq{eq:f_x_expansion} one arrives at 
\begin{eqnarray}
\label{eq:omega_phi_expansion}
\beta\,\omega_{II,0} &\simeq & L^{-3} \xi_0^3x_\tau^3
A \left(-\sign(\tau)\left|\phi\right|^2+\frac{1-M}{2}\; \left|\phi\right|^4+\frac{M}{3}\; \left|\phi\right|^6\right)\\
&=& \left[\frac{(\hbar \Psi_{s,e0})^2}{2m k_B T_c}\xi_0\right]L^{-3} x_\tau^3 \left(-\sign(\tau)\left|\phi\right|^2+\frac{1-M}{2}\; \left|\phi\right|^4+\frac{M}{3}\; \left|\phi\right|^6\right), \nonumber
\end{eqnarray}
where $x_\tau=L/\xi_\tau$, and in the second line we have expressed $A$ in terms of $\xi_0$ and $\Psi_{s,e0}$ on the base of \eq{eq:_corr_lenght}. In this way, it is clear that the effective theory with $d=3,\nu=2/3$ and $\beta=1/3$ is characterized by three parameters---$A, \xi_0$ and $M$---the values of which are determined from  experiment. In \ref{ap:psi_theory} we consider an example of such a theory, the so-called $\Psi$-theory, formulated exactly as in the classical reviews \cite{GS76}and \cite{GS82} by Ginzburg and Sobyanin. As shown there, see \eq{eq:omega2viaphi_ap}, $A=[(3+M)/3]^{1/2}\beta T_{\lambda } \Delta C_{\mu }$ within the $\Psi$-theory. Explicitly, one has
\begin{equation}
\label{eq:omega2viaphi_GS}
\beta \omega_{{\rm II},0}^{\rm GS} = L^{-3} \sqrt{\frac{3+M}{3}} \beta T_{\lambda } {\Delta C}_{\mu }\xi _0^3\; x_{\tau }^3
\left(-\sign(\tau ) \phi ^2+\frac{1}{2} (1-M) \phi ^4+\frac{1}{3}M \phi
^6 \right).
\end{equation}
The value of all constants needed for the evaluation are given in \ref{ap:psi_theory}. In addition to the $\Psi$ theory, we will consider an approach to the $\Psi$-theory where the needed values of the parameters are determined by  experimental measurements of $\xi_0$ and $\Psi_{s,e0}$. In  both approaches the corresponding equation for determining the spatial dependence of $\phi$, see, e.g., \eq{eq:phi_ap}, is given by \eq{eq:phi_final}. Since the behavior of the order parameter is needed for the determination of the Casimir force and because, as stated above, the corresponding equation for the order parameter is the same in both approaches, we start with the determination of the spatial dependence of $\phi$ in a finite film.

\section{The behavior of the order parameter profile and the Casimir force}

\subsection{Analytical results for the behavior of the order parameter profile}
\label{subsec:general_psi}

It is convenient to use the notations 
\begin{equation}
\label{eq:zeta_tau}
\zeta_\tau \equiv \frac{z}{\xi_\tau}=\frac{z}{L}\frac{L}{\xi_\tau}=\zeta_L x_\tau,\qquad \mbox{where} \qquad \zeta_L=\frac{z}{L} \quad \mbox{and} \quad  x_\tau=\frac{L}{\xi_\tau}
\end{equation}
some of which have been introduced in the preceding text but are summarized here for convenience of the reader. 

Multiplying Eq. (\ref{eq:phi_final}) by $\dot{\phi}$ and integrating, we obtain
\begin{equation}
\label{eq:invariant}
\dot{\phi}^2+\sign(\tau) \phi^2 - \frac{1}{2}(1-M) \phi^4 - \frac{1}{3}M\phi^6=p,
\end{equation}
where $p$ is a quantity that is $z$-independent. One should also note that $\dot{\phi}=0$ at $z=L/2$. Let us denote $\phi(z=L/2)=\phi_0$. Then one has 
\begin{equation}
\label{eq:p}
p=\sign(\tau) \phi_0^2 - \frac{1}{2}(1-M) \phi_0^4 - \frac{1}{3}M\phi_0^6.
\end{equation}
Thus, for $\dot{\phi}^2$ one has 
\begin{equation}
\label{eq:phi_dot}
\dot{\phi}^2=\sign(\tau)(\phi_0^2-\phi^2)-\frac{1}{2}(1-M) (\phi_0^4-\phi^4)- \frac{1}{3}M (\phi_0^6-\phi^6).
\end{equation}
At the boundary we have $\phi(0)=0$. This is the minimum value of $\phi$. 
For $T<T_\lambda$, i.e. $\tau>0$, the derivative is greater than zero for $\zeta_\tau$ in the interval  from $\zeta_\tau=0$ to the middle of the system, where it vanishes when the profile levels off close to its bulk value of $\phi_b=1$.  For $T>T_\lambda$, i.e. $\tau<0$, one finds that $\dot{\phi}^2\le 0$ if $\phi(\zeta_\tau)<\phi_0$ for any value of $\zeta_\tau$.  Keeping in mind the fact that $\phi(0)=0$, we conclude that $\phi(\zeta_\tau)=0$ is the only possible real solution in this case.   

Before proceeding to the technical details of the calculations let us note that, according to \eq{eq:first_int}, one has
\begin{equation}\label{p_versus_PL}
p=\frac{1}{2} P_L,
\end{equation}
where $P_L$ is the pressure on the boundaries of a system with size $L$, the behavior of which is mathematically described  by the corresponding functional written in terms of the variable $\phi$. 

The equation for the profile $\phi(\zeta_\tau)$, $\tau>0$, reads
\begin{equation}
\label{eq:profile}
\zeta_\tau=\int_{0}^{\phi (\zeta_\tau)} \frac{d\phi}{\sqrt{(\phi_0^2-\phi^2)-\frac{1}{2}(1-M) (\phi_0^4-\phi^4)- \frac{1}{3}M (\phi_0^6-\phi^6)}},
\end{equation}
complemented by the equation that determines $\phi_0$
\begin{equation}
\label{eq:phi_not}
\frac{1}{2}\frac{L}{\xi_\tau}=\int_{0}^{\phi_0} \frac{d\phi}{\sqrt{(\phi_0^2-\phi^2)-\frac{1}{2}(1-M) (\phi_0^4-\phi^4)- \frac{1}{3}M (\phi_0^6-\phi^6)}}.
\end{equation}
Introducing the variable $\phi=y \phi_0$, and after that performing the change of variables from  $y^2\to y$ the above equation becomes 
\begin{equation}
\label{eq:phi_not_y}
x_\tau \equiv \frac{L}{\xi_\tau}= \int_{0}^{1} \frac{dy}{\sqrt{y\left(1-y\right)\left[1-\frac{1}{2}(1-M)\phi_0^2 \left(1+y\right)- \frac{1}{3}M \phi_0^4\left(1+y^2+y \right)\right]}}.
\end{equation}
The integral on the right-hand side of \eq{eq:phi_not_y}  leads naturally to expressions involving elliptic functions. Furthermore, it is easy to see that when $0\le M\le 1$ the right-hand side of \eq{eq:phi_not_y} is a monotonically increasing function of $0\le\phi_0^2\le 1$. Therefore, when it exists, there is a single solution of \eq{eq:phi_not_y}. Thus, one can uniquely invert this equation, thereby determining  $\phi_0(x_\tau,M)$ for $0\le M\le 1$. Next, since $\phi_0\ge 0$ the minimal value of $x_\tau$ for which such a solution exists is given by the right-hand side of \eq{eq:phi_not_y} with $\phi_0=0$ in it. In this way one obtains $x_\tau(\phi_0=0,M)=\pi$. The last implies that there is a non-zero solution for $\phi_0$ and, therefore, for $\phi(\zeta_\tau,M)$ only for $x_\tau>\pi$. This statement is independent of the value of $M$. Finally, using the above arguments one can determine the position of the minimum value of $p$, and therefore of $P_L$, as a function of $x_\tau$. Indeed, let us note that, for $\tau>0$, $p$ is an increasing function of $\phi_0$, and therefore $x_\tau$ for $0\le \phi_0\le 1$. Thus, the minimal value of $p$ is achieved at the minimal value of $\phi_0$, i.e., at $x_\tau=\pi$, which is also where the Casimir force takes on its largest negative value.  Thus, we come to the conclusion that the position of the largest negative value of the Casimir force does not depend on $M$ within the class of effective theories considered above. Experimental investigations \cite{GSGC2006} yield for the position of the minimum $x_\tau=3.2\pm 0.18$, which effectively coincides with our result. We will see, however,  that the value of the minimum and the shape of the scaling function of the Casimir force depend on $M$. To that end we need more detailed knowledge for the behavior of the order parameter profile.

Below we provide the mathematical derivation of the order parameter profile for different values of $M$. As we will see, the results for $0<M<1$ contain division by $M$. This is why we start with the treatment of the $M=0$ case. Then we will see that if one takes in the expressions for $M\ne 0$ the limit $M\to 0$ one obtains the $M=0$ result.   

\subsubsection{The case $M=0$}
In this case the expression for $p$ becomes 
\begin{equation}
\label{eq:p}
p=\sign(\tau) \phi_0^2 - \frac{1}{2} \phi_0^4,
\end{equation}
while for $x_\tau$, from \eq{eq:phi_not_y}, one has
\begin{equation}
\label{eq:xt}
x_{\tau } = \int_0^1 \frac{1}{\sqrt{y (1-y) \left({\rm sgn}(\tau )-\frac{1}{2} \phi _0^2
		(1+y)\right)}} \, dy.
\end{equation}

From \eq{eq:xt} it is clear that $x_\tau$ is a well defined quantity only for $0\le\phi_0<1$. It is also easy to check that $x_\tau$ is a monotonically increasing function of $\phi_0$, with $x_\tau(\phi_0=0)=\pi$. The last implies that one will have a non-zero solution for $\phi_0$ and, therefore, for $\phi(\zeta_\tau)$ for $x_\tau>\pi$. We note that  \eq{eq:p} leads to $0 \le p<1/2$.  
\begin{figure}[h!]
	\includegraphics[width=\columnwidth]{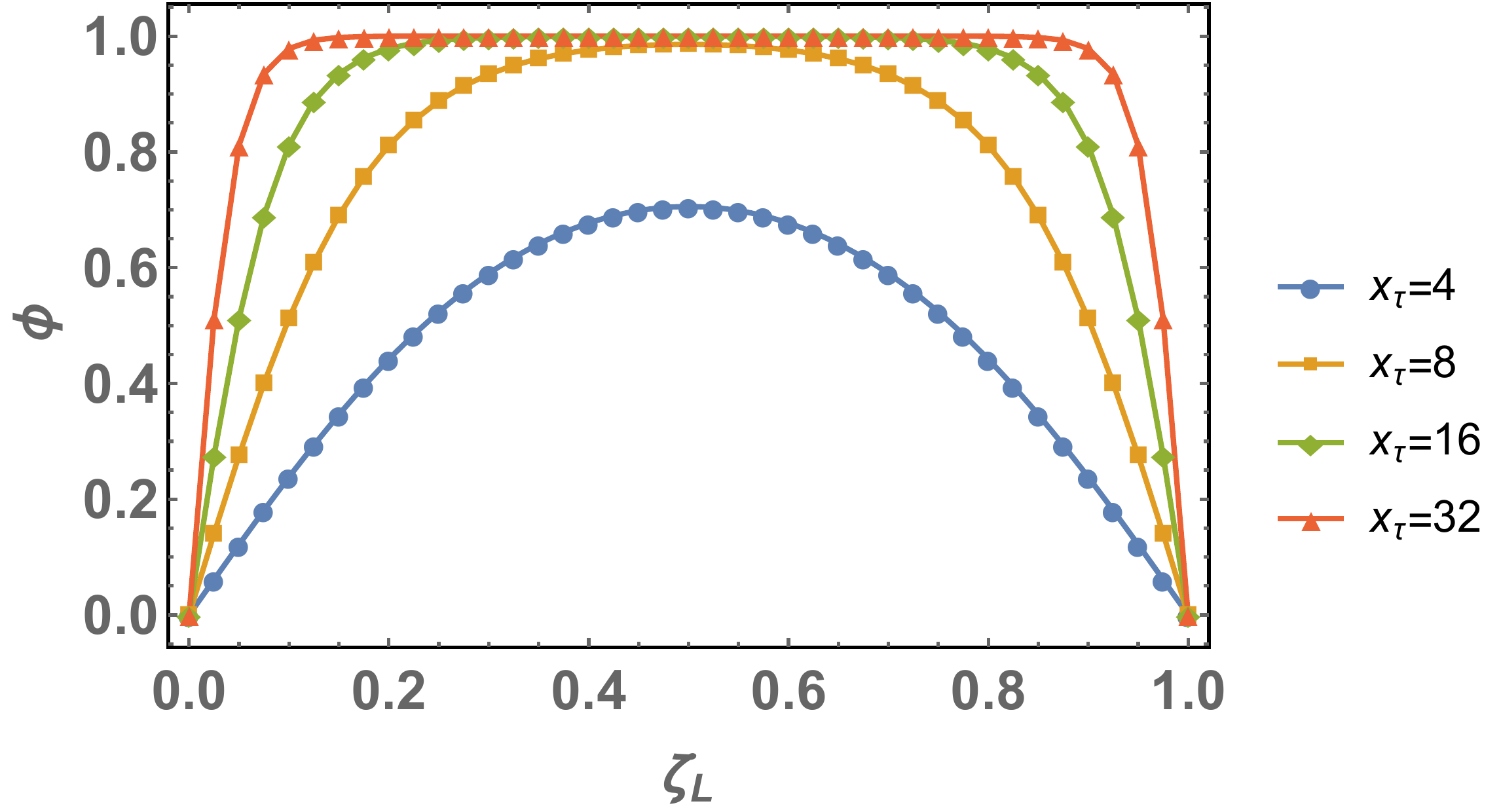}
	\caption{Several profiles of the order parameter for different values of $x_\tau$ for $M=0$ are shown.
	} 
	\label{fig:profiles_M0}
\end{figure}

Taking the integral in \eq{eq:xt}, when $0\le\phi_0<1$ one derives
\begin{eqnarray}
\label{eq:xt_tau_pos}
x_{\tau }= \frac{2 \sqrt{2} }{\sqrt{2-\phi_0^2}}K\left(\sqrt{\frac{\phi _0^2}{2-\phi_0^2}}\right),
\end{eqnarray}
where $K(k)$ is the complete elliptic integral of the first kind of elliptic modulus $k$.
It is easy to check that the right-hand side of \eq{eq:xt_tau_pos} is a monotonically increasing function of $\phi_0$. Thus, one can uniquely invert this equation, thereby determining  $\phi_0(x_\tau)$. 
As noted previously, when  $\phi_0\ge\phi(\zeta_\tau)$,  one is directly led to the conclusion that $\phi(\zeta_\tau)=0$ is the only allowed solution of \eq{eq:phi_dot}.

Solving \eq{eq:profile}, for the order parameter profile $\phi(\zeta_\tau)$ in the case $M=0$ one has
\begin{equation}\label{eq:profileM0}
\phi(\zeta_\tau)=\phi _0 \; {\rm sn}\left(\zeta_\tau  \sqrt{1-\frac{\phi _0^2}{2}}\Bigg|\sqrt{\frac{\phi
		_0^2}{2-\phi _0^2}}\right),
\end{equation}
where $\phi_0$, as a function of $x_\tau$, is to be determined from \eq{eq:xt_tau_pos}. Here ${\rm sn}$ is the Jacobi elliptic function ${\rm sn}(u|m)$ \cite{AS}. The behavior of $\phi(\zeta_L)$, $0\le\zeta_L\le 1$, for several values of $x_\tau$ is shown in Fig. \ref{fig:profiles_M0}.

\subsubsection{The case $0<M<1$}

From \eq{eq:phi_not_y} it is easy to check that $x_\tau$ is a monotonically increasing function of $\phi_0$, with $x_\tau(\phi_0=0, M)=\pi$. The last implies that there will be a unique non-zero  solution for $\phi_0$ and, therefore for $\phi(\zeta_\tau)$, only when $x_\tau>\pi$. The above statements are valid for {\it any} value of $0\le M<1$.

Evaluating the integral in \eq{eq:phi_not_y}, for $\phi_0$ one obtains the equation
\begin{equation}
\label{eq:phi_not_y1}
x_\tau \equiv \frac{L}{\xi_\tau}= \frac{4 \sqrt{3}}{b} K(k) \qquad \mbox{with} \qquad k=\sqrt{ 2 \sqrt{3} a}\; \frac{\phi_0 }{b}
\end{equation} 
and
\begin{equation}
a =\sqrt{(3+M+2M\phi_0^2) (1+M(3-2\phi_0^2)) },\quad 
b =\sqrt{12-6 M\phi_0^4+\phi_0^2 ( \sqrt{3} a-9(1-M))}\,.
\end{equation}
Solving this equation for $\phi_0$, one finds $\phi_0=\phi_0(x_\tau, M)$.

\begin{figure}[h!]
	\includegraphics[width=\columnwidth]{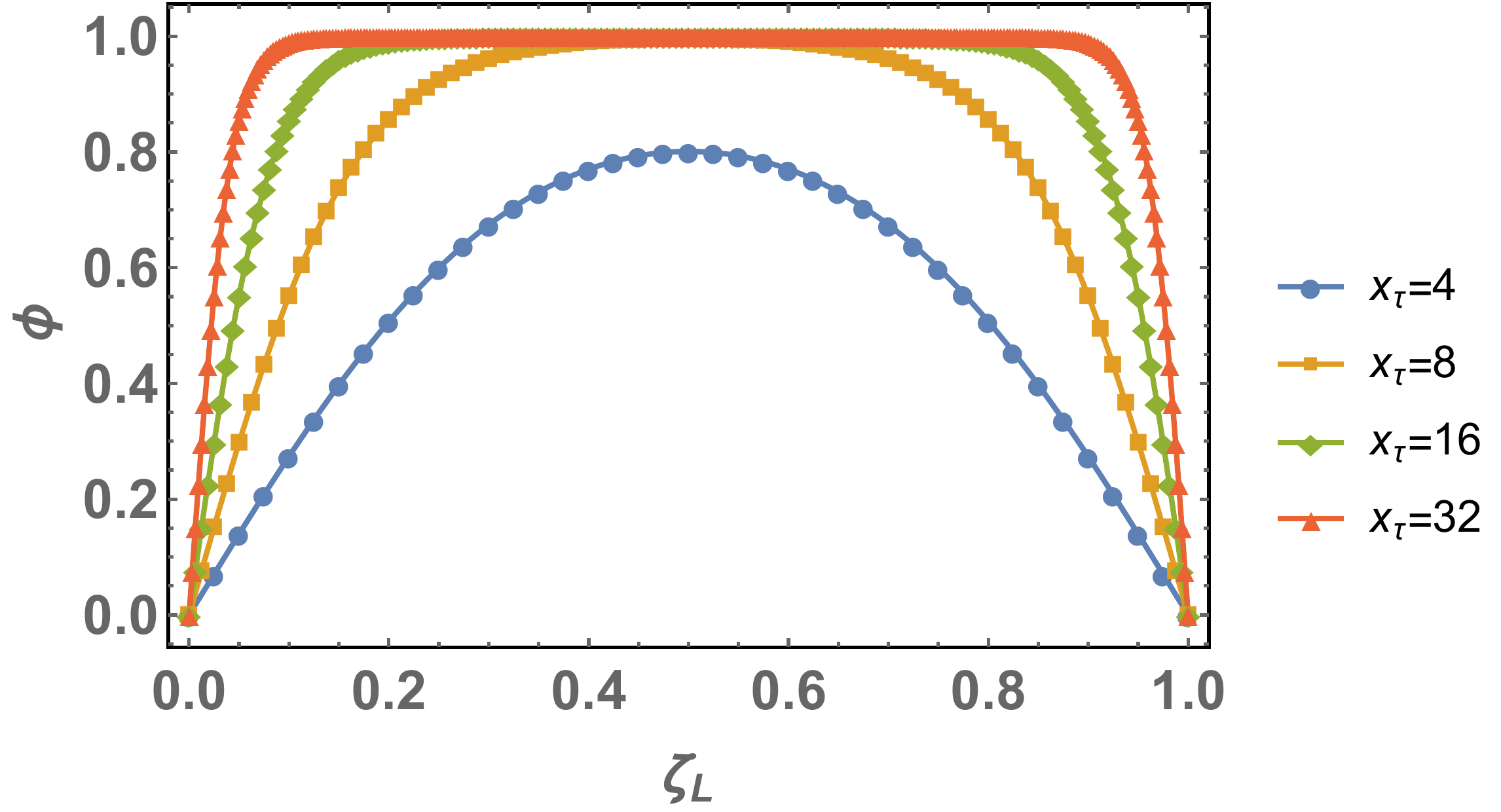}
	\caption{Several profiles of the order parameter $\phi$ for different values of $x_\tau$ for $M=0.5$ are shown. One observes that for the same $x_\tau$ they have larger maximal values of the order parameter than those with $M=0$.
	} 
	\label{fig:profiles_M05}
\end{figure}

The order parameter profile can be also obtained in an explicit form for the case $0\le M<1$. Solving \eq{eq:profile}, for the order parameter profile $\phi(\zeta_\tau|M)$ one has
\begin{equation}\label{eq:profileM_not_0}
\phi(\zeta_\tau|M)=\frac{\sqrt{c} \, \phi _0  \, \mathrm{sn}\left(\frac{b}{2 \sqrt{3}} \zeta_{\tau} | k \right)}{\sqrt{1+c-\mathrm{sn} \left(\frac{b}{2 \sqrt{3}}\, \zeta_{\tau} |k \right)^2}}, \qquad \mbox{where} \qquad 
c=\frac{\sqrt{3} a+M \left(2 \phi_0 ^2 -3\right)+3}{4 M \phi_0 ^2}.
\end{equation}
Here $\phi_0$, as a function of $x_\tau$, is to be determined from \eq{eq:phi_not_y1} and  ${\rm sn}$ again symbolizes the Jacobi elliptic function ${\rm sn}(u|m)$. The behavior of $\phi(\zeta_L,M)$, $0\le\zeta_L\le 1$, for several values of $x_\tau$ and for $M=0.5$ is shown in Fig. \ref{fig:profiles_M05}.

\subsection{The behavior of the Casimir force}

In order to obtain the behavior of the Casimir force, we make use of the first integral \eq{eq:p}, its relation to the pressure in the finite system \eq{p_versus_PL}, the readily obtainable expression for the bulk pressure, as well as the relation \eq{eq:xt} between $\phi_0$ and $x_t$, and finally the corresponding expression for $\beta \omega_{{\rm II},0}$. In this way one derives
\begin{equation}\label{eq:Xcas_general}
\beta F_{\rm Cas}(T,L;M)=Q x_\tau^3 \left[p(\phi_0(x_t,M), M)-\frac{3+M}{6}\right] L^{-3}, 
\end{equation}
where 
\begin{equation}
\label{eq:QGS}
Q=Q^{\rm GS}=\frac{1}{2}\beta T_{\lambda } \Delta C_{\mu } \xi _0^3\;\sqrt{\frac{3+M}{3}}=\sqrt{\frac{3+M}{3}} \left \{\begin{array}{cccl}
0.119, & {\rm with} &\xi_0=1.63 & {\rm as\; reported \; in}\;\mbox{ \cite{GS76,GS82}}\\
0.081, &  {\rm with} &\xi_0=1.432 & {\rm as\; reported\; in}\; \mbox{\cite{TA85}}
\end{array} \right.
\end{equation}
in the case of Ginzburg-Sobyanin theory, and 
\begin{equation}
\label{eq:QGS}
Q=Q^{(\Psi,\,\xi)}=\frac{1}{2}\frac{(\hbar \Psi_{s,e0})^2}{2m k_B T_\lambda}\xi_0=0.106
\end{equation}
in the case of the formulation of the theory in which the constant in $\beta \omega_{{\rm II},0}$ is to be determined via $\Psi_{s,e0}$ and $\xi_0$. Here the value of $\Psi_{s,e0}$ is taken from \cite{GS76,GS82}, while the value of $\xi_0$ is from \cite{TA85}. 

The above implies that for the scaling function of the Casimir force one has
\begin{equation}\label{eq:XCas}
X_{\rm Cas}(x_\tau,M)=Q x_\tau^3 \left[p(\phi_0(x_t,M), M)-\frac{3+M}{6}\right].
\end{equation}
It is clear that the main difference in the both approaches outlined above lies in the way one determines the constant $Q$. This influences the values of the Casimir force, but not the position of the minimum of the force; it is the \textit{same} in both cases, as it has been demonstrated in Sec. \ref{subsec:general_psi}. 

We first establish the value of the minimum of the scaling function of the Casimir force in the context of the model under consideration. We start by recalling that the minimum value of this function  as obtained experimentally \cite{GSGC2006}  is $-1.30$. Since the minimum of $p$ is at $x_\tau=\pi$, and $\phi_0(x_\tau=\pi,M)=0$, from  \eq{eq:XCas} we obtain for the minimum value of the scaling function of the Casimir force 
\begin{equation}\label{eq:XCas_min}
X_{\rm Cas}^{\rm min} \equiv \min_{x_\tau} X_{\rm Cas}(x_\tau,M)=X_{\rm Cas}(x_\tau=\pi,M)=-\frac{3+M}{6} \pi^3 Q.
\end{equation}
Obviously, the minimum deepens as $M$ increases. The results are summarized in the Table below. 
\begin{table}[h]
\label{tab:min_Cas}
\begin{center}
\begin{tabular}{||c|c|c|c|c||}
	\hline 
	\rule[-1ex]{0pt}{2.5ex}  & M=0  & M=0.2 & M=0.5 & M=0.8 \\ 
	\hline 
	\rule[-1ex]{0pt}{2.5ex} Ap1  & -1.845 &- 2.03 & -2.325 & -2.630 \\ 
	\hline 
	\rule[-1ex]{0pt}{2.5ex} Ap2 & -1.256 & -1.383 & -1.582 & -1.790 \\ 
	\hline 
	\rule[-1ex]{0pt}{2.5ex} Ap3 & -1.643  & -1.753 & -1.917  & -2.082 \\ 
	\hline 
\end{tabular}
\end{center}
\caption{The values of the minimum of the scaling function of the Casimir force for different values of $M$. Here with Ap1 we denote the standard Ginzburg-Sobyanin theory with $\xi_0=1.63$, Ap2 - this theory with $\xi_0=1.432$, and with Ap3 - the theory in which the parameter $Q$ is determined via $\Psi_{s,e0}$ and $\xi_0=1.432$.}
\end{table}
The data in this table are not enough to decide which theoretical curve for the scaling function best approximates experimental data.  Before doing that, let us note that $(3+M)/6$ is the value of $p$, see \eq{eq:p}, for $\tau>0$ with $\phi_0=1$. Since $p$ is monotonically increasing function of $0\le \phi_0\le 1$, the last implies that $p<(3+M)/6$ for \textit{any} values of $x_\tau$ and $M$, i.e., 
\begin{equation}
\label{eq:Xcas_neg}
X_{\rm Cas}(x_\tau,M)\le 0.
\end{equation}
The last implies that the Casimir force is \textit{attractive} within the class of theories we are considering. 

\begin{figure}[h!]
	\includegraphics[width=\columnwidth]{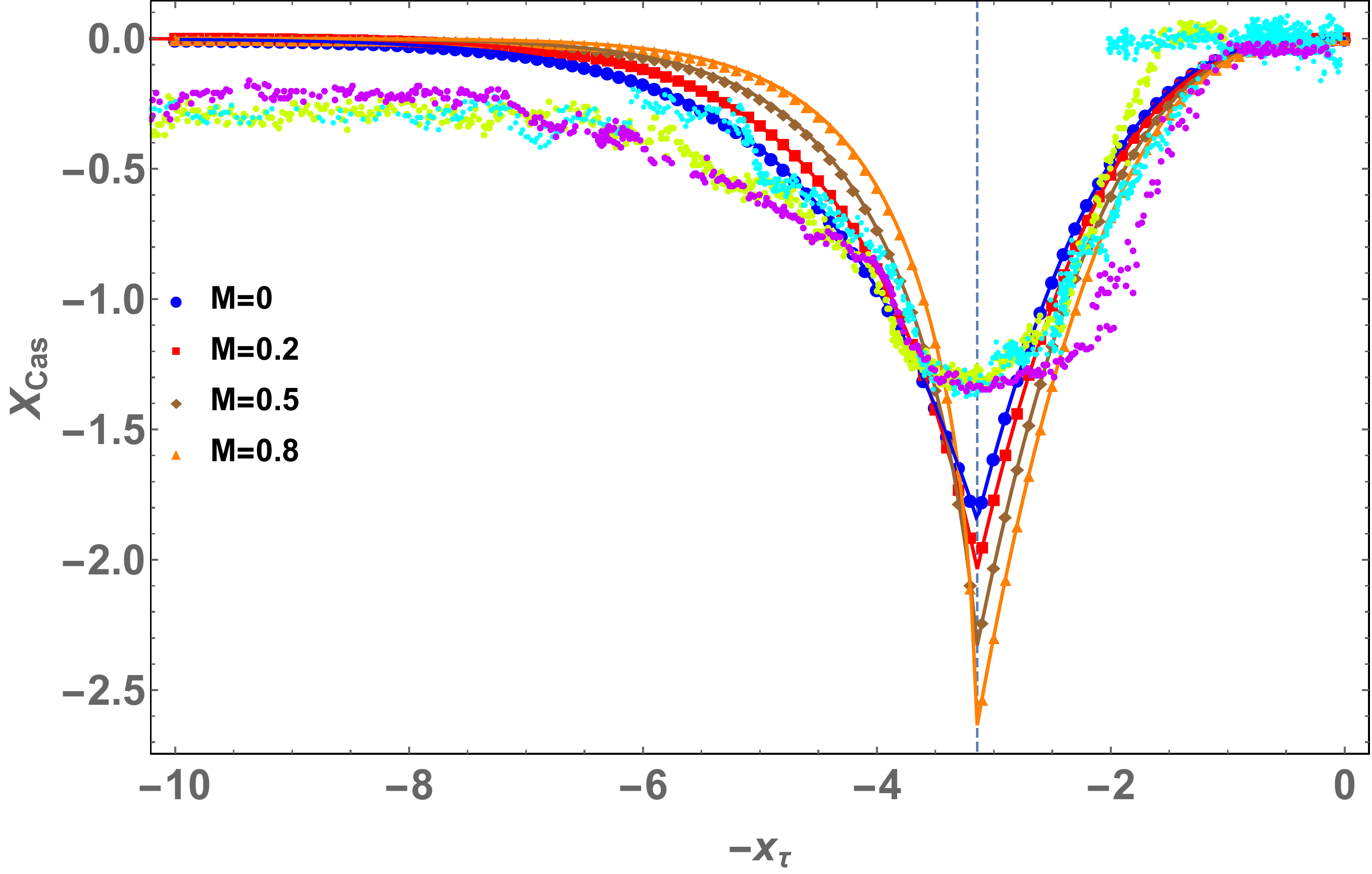}
	\caption{
		A comparison of the experimental data of the Casimir force, the scattered curves,  reported in \cite{GSGC2006}  with the prediction of the $\psi$ theory when $M=0, 0.2, 0.5$, and $M=0.8$, the solid curves. The vertical line marks $x_\tau=\pi$, which is the position of the minimum of the force within the $\Psi$-theory.  It does not depend on $M$. The experiment delivers for the position of the minimum $x_\tau=3.2\pm 0.18$. The minimal value of the force  within the experiment  is $-1.30$. The closest to this is the analytical result with $M=0$ which is $-1.848$. It seems that this curve also goes closest to the experimental data. Due to the large scatter of the experimental data more sophisticated evaluations of the deviations of theoretical curves from the experiment, like the smallest mean square deviation, seems inapplicable.  } 
	\label{fig:PsiSRCF_comp_many_Ms}
\end{figure}

\begin{figure}[h!]
	\includegraphics[width=\columnwidth]{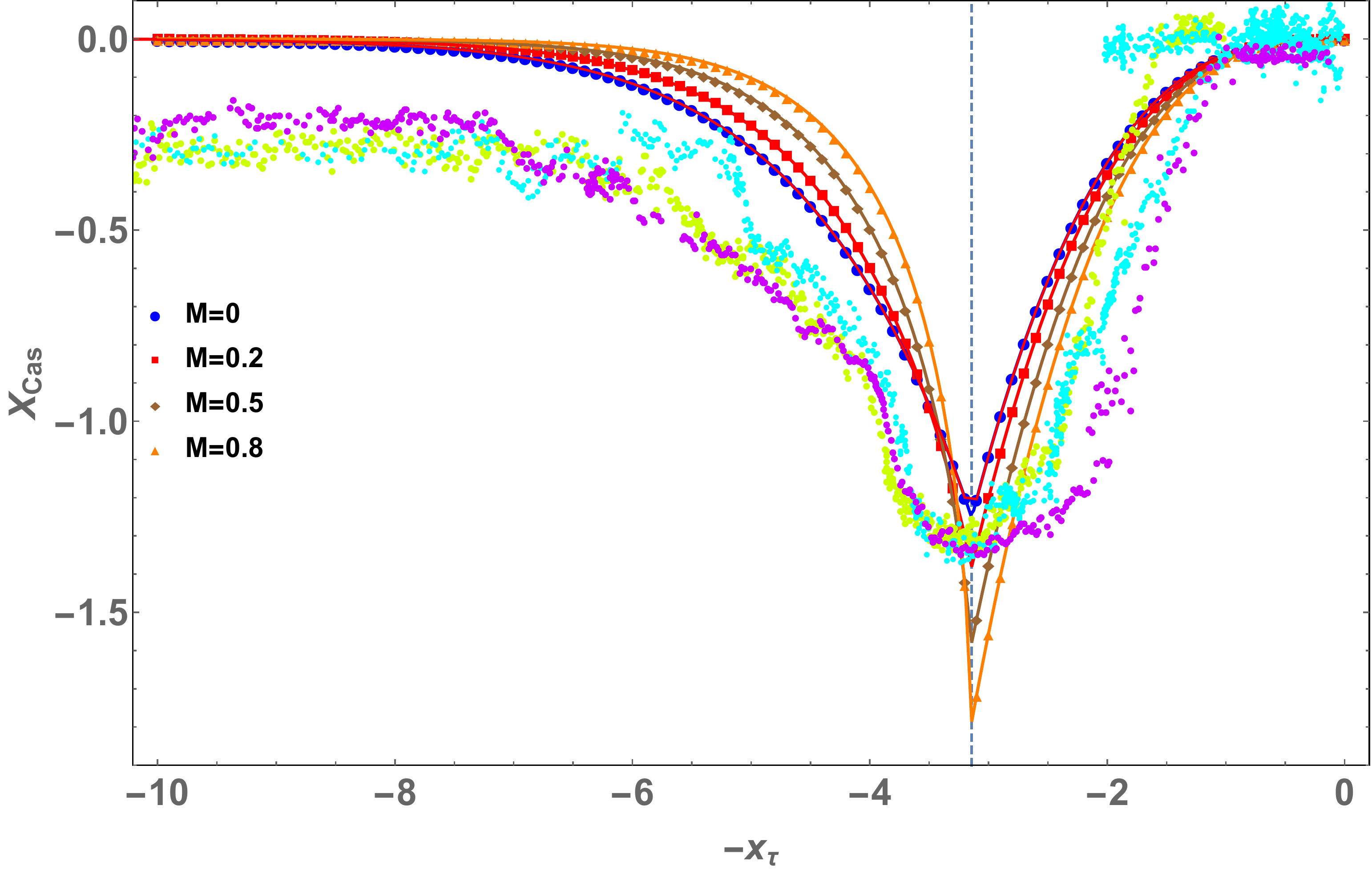}
	\caption{
		A comparison of the experimental data of the Casimir force, the scattered curves,  reported in \cite{GSGC2006}  with the prediction of the \textit{Ap2} approach to $\Psi$-theory when $M=0, 0.2, 0.5$, and $M=0.8$, the solid curves. The vertical line marks $x_\tau=\pi$, which is the position of the minimum of the force within the $\Psi$-theory.  It does not depend on $M$.  } 
	\label{fig:PsiSRCF_comp_many_Ms_xi}
\end{figure}

\begin{figure}[h!]
	\includegraphics[width=\columnwidth]{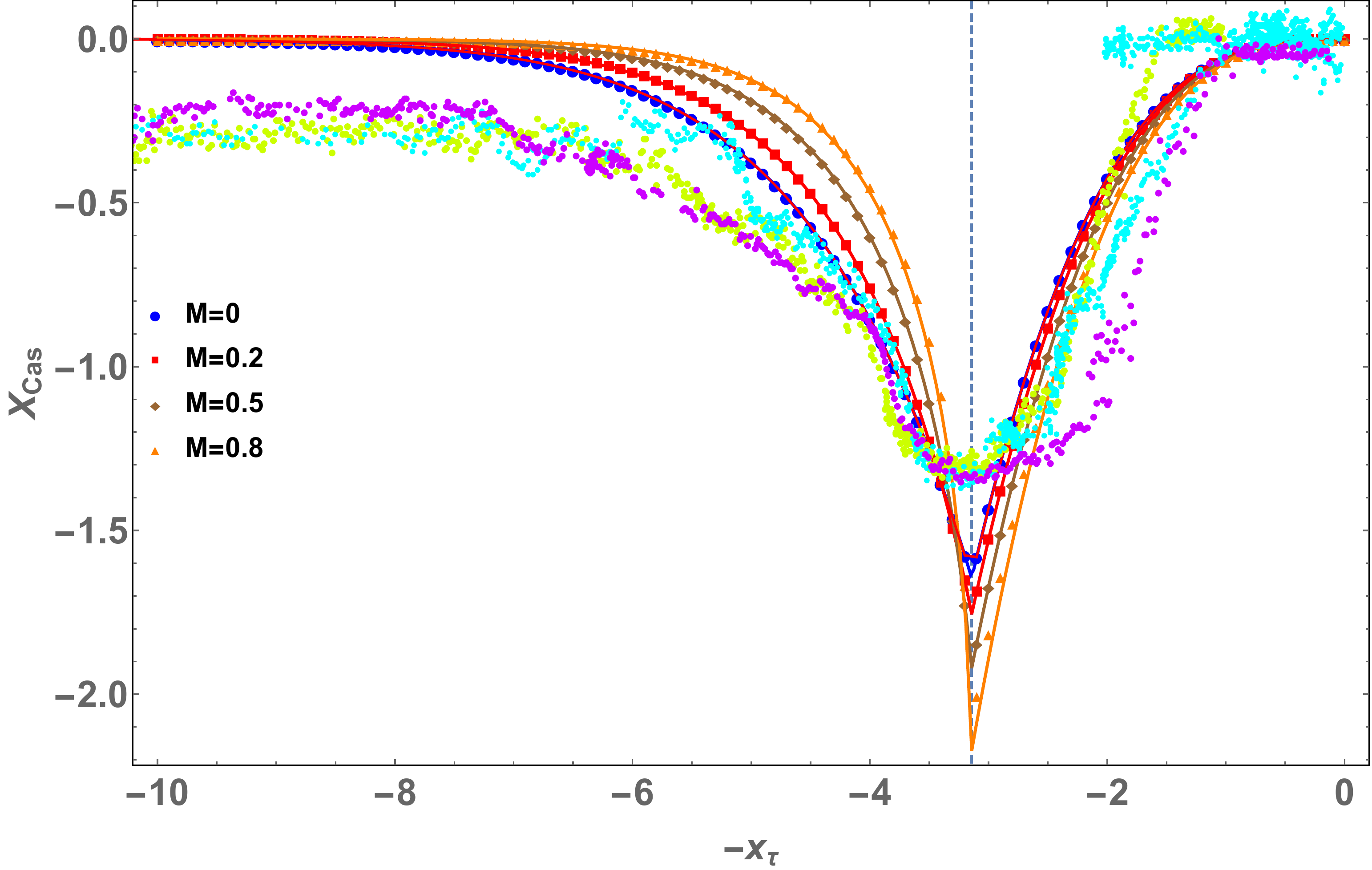}
	\caption{
			A comparison of the experimental data of the Casimir force, the scattered curves,  reported in \cite{GSGC2006}  with the prediction of the \textit{Ap3} approach to $\Psi$-theory when $M=0, 0.2, 0.5$, and $M=0.8$, the solid curves. The vertical line marks $x_\tau=\pi$, which is the position of the minimum of the force within the $\Psi$-theory.  It does not depend on $M$. } 
	\label{fig:PsiSRCF_comp_many_Ms_xi_psi}
\end{figure}

For convenience, let us introduce the following short-hand notation for the approaches considered in this article:  \textit{i)} \textit{Ap1} - the $\Psi$-theory as formulated in \cite{GS76,GS82};  \textit{i)} \textit{Ap2} - the same theory but with the value of the $\xi_0$ amplitude as determined in \cite{TA85};  \textit{iii)} \textit{Ap3} - theory, in which the constant $Q$ is determined via $\Psi_{s,e0}$ and $\xi_0$ from Ref. \cite{TA85}. The scaling function of the Casimir force in any of these approaches are plotted for $M=0, 0.2, 0.5$, and $M=0.8$ in Figs. \ref{fig:PsiSRCF_comp_many_Ms}, \ref{fig:PsiSRCF_comp_many_Ms_xi} and 
	\ref{fig:PsiSRCF_comp_many_Ms_xi_psi} respectively. 
 As we see, the position of the minimum is at $x_\tau=\pi$, independent of the value of $M$, while experiment yields  $x_\tau=3.2\pm 0.18$, which is effectively consistent with our result.  For \textit{Ap1}, the absolute maximum value of the (negative) force is attained at $M=0$; it is $-1.848$ while experimentally it is $-1.30$. Inspection of the plot shows that closest agreement between theory and experiment is obtained for $M=0$. The next best curve is that one with $M=0.2$ for which the minimal value of the force is not too different from the $M=0$ case. One has $X_{\rm Cas}^{\rm min}(x_\tau=\pi,M=0.2)=-2.036$. Attempts have been undertaken to determine the value of the parameter $M$ from experiment \cite{PS82,PS86,SS87,SS88}. The issue has been discussed in \cite{GS87,GS87b,BGS88}. Summarizing the findings, on the whole the experiments analyzed yield  values for the parameter $M$ in the range $0.5\pm 0.3$. If one wants to maintain agreement with this results one is led to choose the value $M=0.2$ from the above set of parameters considered. 	Fig. \ref{fig:PsiSRCF_comp_many_Ms_xi} 	presents a comparison of the experimental data of the Casimir force---the scattered curves,  reported in \cite{GSGC2006}  with the prediction of the \textit{Ap2} theory when $M=0, 0.2, 0.5$, and $M=0.8$, shown as solid curves. Despite the fact that one obtains values for the minimum closer to the experimentally observed ones like $X_{\rm Cas}^{\rm min}(x_\tau=\pi,M=0)=-1.256$ and $X_{\rm Cas}^{\rm min}(x_\tau=\pi,M=0.2)=-1.383$, the overall agreement of the scaling functions in these cases becomes worse than the one predicted for the same values of $M$ in \textit{Ap1}.   	Finally, Fig. \ref{fig:PsiSRCF_comp_many_Ms_xi} 	presents a comparison of the experimental data of the Casimir force, the scattered curves,  reported in \cite{GSGC2006},  with the prediction of the \textit{Ap3} approach to $\Psi$-theory when $M=0, 0.2, 0.5$, and $M=0.8$, the solid curves. One observes a relatively good agreement between the analytical curves and the experimental ones for $M=0$ and $M=0.2$. 
 
\begin{figure}[h!]
	\includegraphics[width=\columnwidth]{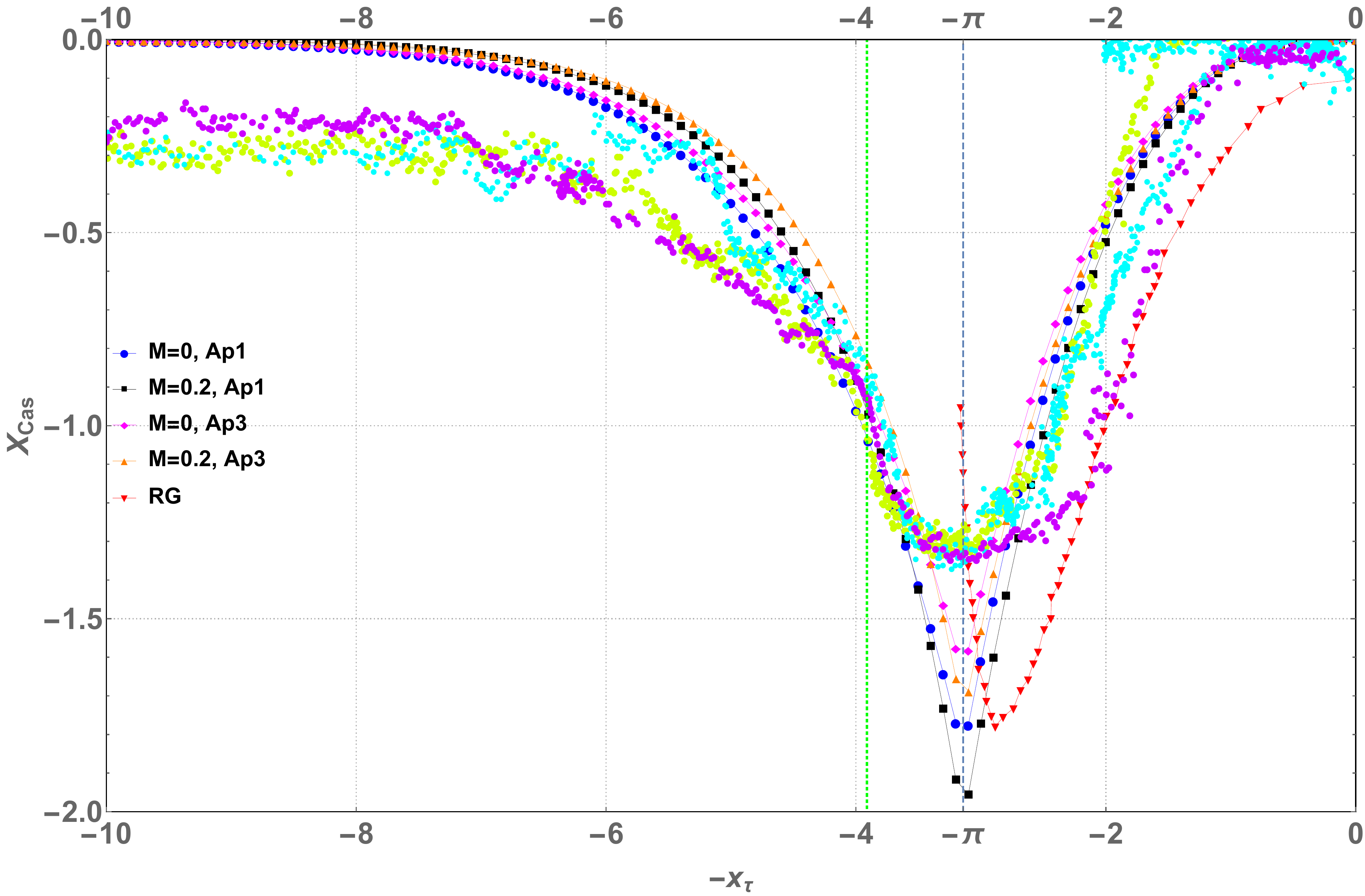}
	\caption{
		A comparison of the experimental data of the Casimir force, the scattered curves,  reported in \cite{GSGC2006}  with the prediction of the $\Psi$ theory when $M=0$ and $M=0.2$, the solid curves. The prediction of the Renormalization Group theory \cite{D2014} is also presented - the red curve with the inverted triangles marks.  The green dashed vertical line at $-3.91$ marks the Monte Carlo prediction for the occurrence of Kosterlitz-Thouless transition in the film \cite{H2009}. Due to the serious scatter of the experimental data it is not possible to quantify which of the four analytical curves presented best fit the experimental data.} 
	\label{fig:PsiSRCF_comp}
\end{figure}

  Figure~\ref{fig:PsiSRCF_comp} shows  the comparison between experiment and the $M=0$ and $M=0.2$ curves for \textit{Ap1} and \textit{Ap2}. We observe two branches of the both sides of $x_\tau=\pi$, for $x_\tau \lesssim 2.5$ and $6\gtrsim x_\tau\gtrsim 3.7$, with very nice agreement between the theoretical curve and  experimental data. The disagreement is\textit{ i)} in the cusp-like region of the theoretical curve very close to $x_\tau=\pi$, which reflects the fact that critical fluctuations are not completely taken into account in the versions of the $\Psi$-theory considered here, and \textit{ii)} for $x_\tau \gtrsim 6$, where Goldstone modes, which provide the dominant contributions,  are also absent in \textit{Ap1} and \textit{Ap2}. We find the overall agreement between  experiment and the results of the theory to be reasonable in light of its evident drawbacks. 

To be precise, let us stress that when transferring the experimental data of \cite{GSGC2006}, given in terms of $(T/T_\lambda-1)L^{1/\nu}$ to the variable $L/(\xi_0 |t|^{1/\nu})$ we have used the value of $\nu$ given in \eq{eq:crit_exp}, and the data of $\xi_0=1.432$ $\AA$ reported in \cite{TA85}.  Figure~\ref{fig:PsiSRCF_comp} also contains some results obtained within different theoretical approaches and Monte Carlo simulations. They will be discussed in the next section. 

\section{Discussion and concluding remarks}

In the study reported here we have applied three variants, which we label \textit{Ap1},  \textit{Ap2}, and \textit{Ap3}, of the $\Psi$-theory of Ginzburg and co-authors \cite{GS76,GS82}, to evaluate the temperature behavior of the order parameter profile and the Casimir force in $^4$He film in equilibrium with its vapor. We have obtained exact closed form expressions for the profiles---see \eq{eq:profileM0} and \eq{eq:profileM_not_0}---and for the force within this theory; see \eq{eq:Xcas_general}. We have found closest agreement between the theory and experiment for $M=0$ and $M=0.2$ for \textit{Ap1} and  \textit{Ap3} which differ from each other by how the parameters of the theory are determined from the experiment. The scaling functions of the Casimir force in any of these versions of the $\Psi$-theory and the comparison with the experiment are presented for $M=0, 0.2, 0.5$, and $M=0.8$ in Figs. \ref{fig:PsiSRCF_comp_many_Ms}, \ref{fig:PsiSRCF_comp_many_Ms_xi} and 
\ref{fig:PsiSRCF_comp_many_Ms_xi_psi}, correspondingly. 
We conclude that there is reasonably good agreement between this theory with $M=0$ and $M=0.2$ within \textit{Ap1} and  \textit{Ap3}, and the experiment; see Fig. \ref{fig:PsiSRCF_comp}. One should note, however, some important differences. In the $\Psi$-theory there is a sharp two-dimensional phase transition with long-ranged order below the critical temperature of the finite system, while in the Helium system one expects a Kosterlitz-Thouless type transition. This feature  is not captured by the $\Psi$-theory. Also missing are the Goldstone modes and surface wave contributions that appear at low temperatures \cite{ZRK2004}. The overall agreement between the result of the $\Psi$-theory and the experiment, shown in Fig. 	\ref{fig:PsiSRCF_comp}  is, however, much better than is provided by mean-field theory \cite{ZSRKC2007}. It is true, however, that in the process of defining the $\Psi$-theory its principal parameters have been determined from experiment. 

Let us now comment on some other problems related to the original formulation of the $\Psi$-theory, which we called \textit{Ap1}. The value 
$\alpha=0$ the $\Psi$-theory is consistent with  logarithmic type dependence of the specific heat. This is, of course, an approximation to the behavior of $^4$He, since high precision experimental measurements yield a slightly negative value of $\alpha$---see \eq{eq:universality}---which tells us that the specific heat is finite and continuous,  with a cusp-like singularity at $T_\lambda$. Thus $\Delta C_\mu$ used in the \textit{Ap1} version of the $\Psi$ theory should also be considered  an approximation of the experimental situation. The corresponding problem is clarified below but before commenting on it, let us mention that \textit{Ap3} version of the theory actually lacks such a problem.  In the logarithmic case one usually writes 
\begin{equation}
\label{eq:C-mu}
C_\mu(\tau)=\left \{ 
\begin{array}{ll}
A \ln|\tau|+B, & \tau<0\\
A \ln \tau+B+\Delta C_\mu, & t>0
\end{array}.
\right.
\end{equation}
In the general case of small $\alpha$ one writes the specific heat behavior in the form  \cite{GS82,P90,PHA91,SD2003,BHLD2007}
\begin{eqnarray}
\label{eq:spec_heat}
\lefteqn{C(\tau) \simeq \frac{A^{\pm}}{\alpha}(|\tau|^{-\alpha}-1)+\cdots+B},\\
&\approx &-A^{\pm} \ln|\tau| \left\{1-\frac{1}{2}\alpha \ln|\tau|+\mbox{{\cal O}}\left[(\alpha\ln|t|)^2\right] \right\}+\cdots+B,\nonumber 
\end{eqnarray}
where $\cdots$ refers to corrections to scaling. In order for the second row of the above expressions to make sense one must assume values of $\tau$ for which $|\alpha \ln |\tau||\ll 1$. 

The $\Psi$-theory has problems with universality. It is well known that the scaling functions of the free energy and of the Casimir force are universal. From \eq{eq:Xcas_general} and \eq{eq:XCas} the last implies that $Q$ and $M$ are universal quantities.  In \eq{eq:spec_heat}
\begin{equation}
\label{eq:Rpm_univ}
R^{\pm}=\left(A^{\pm}\right)^{1/3}\xi_0^{\pm}
\end{equation}
is a universal quantity \cite{PHA91,SD2003}. 
It is clear that if one considers $\Delta C_{\mu } \xi _0^3$ as an approximation to an universal number, in line with  \eq{eq:Rpm_univ} which provides an universal number, then the expression for the Casimir force is also ``universal'' within the same level of approximation. A determination of $M$ from various experiments actually shows that $M$ is \textit{not} a universal quantity, which of course leads to the absence of universality within the $\Psi$-theory.

Let us now briefly comment of other existing analytical approaches to the behavior of the Casimir force in $^4$He film. We start with the mean-field theory. 

The $\Psi$-theory is conceptually similar to mean-field theory in that an effective free energy is minimized, without taking fully the fluctuations into account. Within the $\Psi$-theory the role of fluctuations is partially reflected by the choice of the critical exponent $\nu=2/3$. Furthermore, the   $\Psi$-theory is an effective $d=3$ theory, while mean-field theory characteristically applies when $d\ge 4$. Mean-field theory involves a single non-universal parameter, the value of which is determined from information outside this theory in order to make a connection with experimental data. In Ref. \cite{ZSRKC2007}  Renormalization Group arguments are utilized to determine this nonuniversal parameter.  The minimum of the force in the mean-field case is at $x_{{\rm min, MF}}=\tau(L/\xi_0 )^{1/\nu_{{\rm MF}}} =\pi^2$, where $\nu_{{\rm  MF}}=1/2$. When Renormalization Group input is utilized to ``tune'' mean-field theory as in Refs. \cite{ZSRKC2007} and \cite{MGD2007} the minimum of the scaling function of the force is found to be $X_{\rm Cas, min, MF}=-6.92$
 at $x_{{\rm min, MF}}=\pi^2$. Note, however, that the mean-field scaling variable can be redefined in the form $\tau^{\nu_{\rm MF}} (L/\xi_0 )$. Then the minimum of the force is at   $\hat{x}_{{\rm \tau,min}}^{{\rm MF}}=\sqrt{\tau} (L/\xi_0 )=\pi$. In the $\Psi$-theory one also has minimum at $x_{{\rm \tau,min}}=\tau^{3/2} (L/\xi_0 )=\pi$. Thus, if one writes the scaling variables in the both theories in terms of $L$ divided by the corresponding correlation length, i.e. in terms of $L/\xi_{{\rm MF}}$ or $L/\xi_\Psi$,  both theories will have a minimum at $\pi$. The agreement with the experimental data is, however, better with the $\Psi$-theory, because the experimental value of $\nu$, see \eq{eq:crit_exp}, is quite close to the value $\nu=2/3$ of the $\Psi$-theory. 
 
 The Renormalization Group approach is, of course, the only one that does not need any additional input, as long as one insists on a full validity of the universality hypothesis.  The first attempt in that direction was made in \cite{KD92a,KD92b}, where the authors studied within the $\varepsilon$ expansion the behavior of the force in $^4$He above $T_\lambda$. Recent progress below $T_\lambda$ has been made in \cite{D2014};  see the red curve marked with inverted triangles in Fig. \ref{fig:PsiSRCF_comp}. The proposed theory holds only for temperatures above some temperature $T_{\rm c,film}$, close under $T_\lambda$, and breaks below $T_{\rm c,film}$. Nevertheless, this temperature interval encompasses the position of the minimum of the force. It is reported to be at $x_{\rm min, RG}=\tau_{\rm min, RG}(L/\xi_0 )^{1/\nu}=-4.73$. If one recalculates this in terms of $x_\tau$, one obtains that the minimum is at $\hat{x}_{{\rm \tau,min}}^{{\rm RG}}=2.84$. The value of the scaling function of the Casimir force at the minimum  is not reported in Ref. \cite{D2014}, but from the presented plot, see Fig. $1(b)$, it can be evaluated to be about $X_{{\rm Cas, min, RG}} \simeq -1.8$, which is very close to our finding within \textit{Ap1} with $M=0$. 

Given the criticism on the $\Psi$-theory summarized above, it is nevertheless worthwhile to recall its advantages. It is relatively simple, and despite its many drawbacks it provides analytical results reasonably close to those observed experimentally; see Fig. \ref{fig:PsiSRCF_comp}. That is why we believe that it can be a useful tool for obtaining approximate data for the experimental behavior of a system as interesting and non-trivial as Helium films. This is especially true if one takes into account the state of the art of the more advanced theories, such as those based on the Renormalization Group, as applied to the phenomena considered here especially in the range of temperatures below the critical one. Again refer to Fig. \ref{fig:PsiSRCF_comp}. 

\appendix 

\section{A realization of the effective field theory within the so-called $\Psi$ theory} 
\label{ap:psi_theory}

In accord with \cite{GS82,GS76}, we take $\omega_{II,0}$ to be of the form
\begin{eqnarray}
\label{eq:II_pot}
\omega_{II,0}(\mu,T,|\Psi_s|^2)=\frac{3T_\lambda \Delta C_\mu}{3+M}\left[-\tau |\tau|^{1/3} \left|\frac{\Psi_s}{\Psi_{s,e0}}\right |^2  + \frac{1-M}{2}|\tau|^{2/3}\left|\frac{\Psi_s}{\Psi_{s,e0}}\right |^4 + \frac{M}{3}\left|\frac{\Psi_s}{\Psi_{s,e0}}\right |^6 \right], 
\end{eqnarray}
where $T_\lambda(\mu)$ is the $\lambda$-transition temperature in
equilibrium with saturated vapor, $T_\lambda(\rho_\lambda) = 2.172$ K, $\rho_\lambda = 0.146$ g cm$^{-3}$. Here
\begin{equation}
\label{eq:tau}
\tau=\left[T_\lambda(\mu)-T\right]/T_\lambda(\mu),
\end{equation}
$\Delta C_\mu $ is the specific heat jump
at the $\lambda$ point $\Delta C_\mu = \Delta C_p = 0.76 \times 10^7$ erg cm$^{-3}$ K$^{-1}$, $M$ is a real number serving as a parameter of
the theory, and $\Psi_{s,e0}$ is the amplitude of the temperature dependence of
the equilibrium value of $\Psi_s$ in bulk Helium,
\begin{equation}
\label{eq:psi_eq_value}
\Psi_{s,e}(\tau) =\Psi_{s,e0}\, \tau^\beta = 0.23 \times 
10^{12} \, \tau^{1/3} \, {\rm cm}^{-3/2}.
\end{equation}
The value of $\Psi_{s,e0}$ is, as usual \cite{GS82}, determined by the equation 
\begin{equation}
\label{eq:bulk_psi}
\left(\frac{\partial \omega_{II,0}}{\partial |\Psi|^2}\right)_{\mu,T}=0
\end{equation}
so as to be in accord with the experimental data \cite{GS76,GS82}
\begin{equation}
\label{eq:bul_value_Psi}
\rho_{se}= m|\Psi_{s,e}|^2=0.35\,\tau^{2/3}\; {\rm g}\, {\rm cm}^{-3} =\rho_{s0}\, \tau^\zeta
\end{equation}
with $\zeta\simeq 2\beta\simeq 2/3$.

As is clear from \eq{eq:II_pot}, it is convenient to introduce the reduced variable
\begin{equation}
\label{eq:small_psi}
\psi=\frac{\Psi_s}{\Psi_{s,e0}}.
\end{equation}
Then, \eq{eq:II_pot} becomes 
\begin{equation}
\label{eq:II_pot_small_psi}
\omega_{II,0}=\frac{3T_\lambda \Delta C_\mu}{3+M} \left[-\tau |\tau|^{1/3} \left|\psi\right|^2 + \frac{1-M}{2} |\tau|^{2/3} \left|\psi\right|^4 + \frac{M}{3}\left|\psi\right|^6 \right].
\end{equation}

The above expressions for $\omega$ and $\omega_{II}$ are consistent with a close approximation to the critical exponents in which $\alpha=0$, $\nu=2/3$ and the anomalous dimension exponent $\eta$ is zero.  They define an effective $3$-dimensional theory for the behavior of the Helium films. The best known values of the critical exponents $\alpha$ and $\nu$ for Helium are given above, in \eq{eq:crit_exp}.

Within the $\Psi$ theory the type of phase transition in Helium films from
Helium I to Helium II depends crucially on the value of the parameter $M$ \cite{GS76,GS82}. For $M < 1$ this transition, as in bulk Helium, is continuous, while
for $M > 1$ the transition in a film is first order. The value
$M = 1$ corresponds to a tricritical point. Thus, we use $M<1$ in our calculations. Obviously, the simplest case has $M=0$.

Introducing, as in \cite{GS76,GS82}, the scaled spatial variable
\begin{equation}
\label{eq:zeta_def}
\zeta_0= z/\xi_0,
\end{equation}
where, see Eq. (23) in \cite{GS82}, for $\xi_0$ one has
\begin{equation}
\label{eq:xi0}
\xi_0=\frac{\hbar \Psi_{s,e0}}{\sqrt{2m T_\lambda \Delta C_\mu}}=\frac{\hbar}{m}\sqrt{\frac{\rho_{s,0}}{2 T_\lambda \Delta C_\mu}}\simeq 1.63 \times 10^{-8} {\rm cm}
\end{equation}
with $\xi_0$ being the amplitude of the correlation function above the $\lambda$ point for the version of the theory
with $M = 0$, one can write the equation for the dimensionless function $\psi$ in the form 
\begin{equation}
\label{eq:psi_red_eq_expl}
\ddot{\psi}=\frac{3}{3+M} \psi \left[-\tau |\tau|^{1/3} + (1-M) |\tau|^{2/3} \left|\psi\right|^2 + M\left|\psi\right|^4 \right].
\end{equation}
Here the differentiation is to be understood with respect to the scaled variable $\zeta_0$. \eq{eq:psi_red_eq_expl} is the main equation within the $\Psi$ theory one deals with.

The proper boundary conditions are
\begin{equation}
\label{eq:bc_spec}
\psi(0)=0, \psi(L)=0,
\end{equation}
but so that $\lim_{L\to \infty} \psi(L/2)=\psi(\infty)=\psi_e=\tau^{1/3}$.

It is convenient to introduce the variables
\begin{equation}
\label{eq:xi_t_psi_theory}
\xi_\tau=\sqrt{\frac{3+M}{3}}\,\xi_0\, |\tau|^{-2/3}, \quad \mbox{and} \quad \phi=\psi |\tau|^{-1/3},
\end{equation}
where $\xi_0$ is given by \eq{eq:xi0}. Then \eq{eq:psi_red_eq_expl} becomes 
\begin{equation}
\label{eq:phi_ap}
\ddot{\phi}=\phi \left[-\sign(\tau) + (1-M) \left|\phi\right|^2 + M\left|\phi\right|^4 \right],
\end{equation}
where the derivative is taken with respect to $\zeta_\tau=z/\xi_\tau$.
Note that since $\xi_\tau$ depends on $M$, see \eq{eq:xi_t_psi_theory}, the scaling variable $x_\tau$---see \eq{eq:zeta_tau}---is also $M$-dependent. Note also that, in contrast to commonly utilized notations, $\tau>0$ corresponds to $T<T_\lambda$. Obviously, in equilibrium bulk Helium, when $\ddot \phi=0$, one has  $\phi \equiv \phi_b$ with $\phi_b=1$ for $T\le T_\lambda$, and $\phi_b=0$ for $T>T_\lambda$. In terms of $\phi$ and $x_\tau$,  \eq{eq:II_pot} for $\omega _{{\rm II},0}$ becomes 
\begin{equation}
\label{eq:omega2viaphi_ap}
\beta \omega_{{\rm II},0} = L^{-3} \sqrt{\frac{3+M}{3}} \beta T_{\lambda } {\Delta C}_{\mu }\xi _0^3\; x_{\tau }^3
\left(-\sign(\tau ) \phi ^2+\frac{1}{2} (1-M) \phi ^4+\frac{1}{3}M \phi
^6 \right).
\end{equation}
Note that the above expression is in a full conformity with \eq{eq:f_x_expansion}.

\section*{ACKNOWLEDGMENTS}

We are indebted to the authors of \cite{GC99} and \cite{GSGC2006} for providing their experimental data in electronic form. D. D., V. V. and P. D. gratefully acknowledge the  financial support via contract DN02/8 of Bulgarian NSF. J. R. is pleased to acknowledge support from the NSF through DMR Grant No. 1006128.

\end{document}